\documentclass[aps,prl,showpacs,onecolumn]{revtex4}
\usepackage{graphicx}
\usepackage[usenames]{color}

\def\bea{\begin{eqnarray}}
\def\eea{\end{eqnarray}}
\def\be{\begin{equation}}
\def\ee{\end{equation}}

\def\beq{\begin{equation}}
\def\eeq{\end{equation}}

\begin{document}
\title{\bf Amplification of High Harmonics Using Weak Perturbative High Frequency
Radiation}

\author{Avner Fleischer and Nimrod Moiseyev}

\affiliation{Schulich Faculty of Chemistry and Minerva Center for
Nonlinear Physics of Complex Systems, Technion -- Israel Institute
of Technology, Haifa 32000,
Israel.}\email{avnerf@tx.technion.ac.il ,
nimrod@tx.technion.ac.il}

\date{\today}
\begin{abstract}
The mechanism underlying the substantial amplification of the
high-order harmonics $q \pm 2K$ ($K$ integer) upon the addition of
a weak seed XUV field of harmonic frequency $q\omega$ to a strong
IR field of frequency $\omega$ is analyzed in the framework of the quantum-mechanical Floquet formalism and the semiclassical re-collision model. According to the Floquet analysis, the high-frequency field induces transitions between several Floquet states and leads to the appearance of new dipole cross terms. The semiclassical
re-collision model suggests that the origin of the enhancement lies in the time-dependent modulation of the ground electronic state induced by the XUV
field.

\end{abstract}
\maketitle

{03.65.-w, 42.50.Hz, 42.65.-Ky, 32.80.Rm}

Focusing intense linearly-polarized monochromatic IR laser pulses
into gas of atoms can lead to the emission of high-energy photons
with frequencies extending into the extreme ultraviolet (XUV) and
X-ray region by high harmonic generation (HHG). The HHG phenomena
stands as one of the most promising methods of producing short
attosecond pulses (as-pulses) \cite{Paul Science}.


The contamination of the strong IR field with a second \cite{T. Pfeifer,T. Pfeifer+L. Gallmann,N. Dudovich} or more \cite{Markus
Kitzler,Enrique Conejero Jarque,M. B. Gaarde}
weak XUV fields has a dramatic effect on the dynamical behavior of
the electrons, and had drawn a lot of attention in recent years.
On the basis of the three-step (re-collision) model \cite{P.
B. Corkum,M. Lewenstein,K. J. Schafer} it had been argued that the
role of the XUV field is to switch the initial step in the
generation of high harmonics from tunnel ionization to the more
efficient single XUV-photon ionization. This might explain the
improved macroscopic HHG signal obtained in experiments: the
XUV-assisted ionization increases the number of atoms which
participate in the HHG process and improves phase matching
\cite{A. Heinrich}.

\begin{figure}[ht]
\hbox{\includegraphics[width=2.8in,height=2.8in]{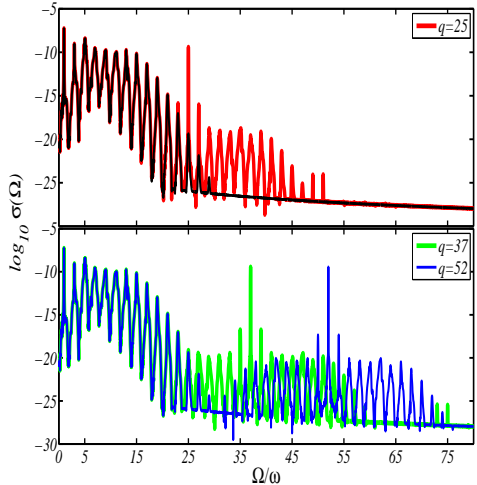}}
\caption{\label{fig1}\small (color online) HGS obtained from a 1D
model Hamiltonian of Xe atom irradiated by a 50-oscillation
sine-square pulse of bichromatic laser field composed of a
\textit{strong} laser field of frequency $\omega$
($\lambda=800nm$) and amplitude $\varepsilon^{in}_{1}$
($I^{in}_{1}\simeq 4.3\cdot 10^{13}W/cm^{2}$) and a \textit{weak}
field of frequency $q\omega$ and amplitude $\varepsilon^{in}_{q}$
($I^{in}_{q}\simeq 3.5\cdot 10^{8}W/cm^{2}$) for different values
of $q$: $q=25$ (solid red line), $q=37$ (dotted green line),
$q=52$ (solid blue line). HGS in the absence of the XUV field is
shown in the dotted black line where the position of the cutoff is
at the 15th harmonic. The harmonics above the
29th harmonic, are enhanced in the addition of the XUV field,
despite of its small intensity. In addition, with respect to its
center $q$, the distribution of the new harmonics in the HGS is
symmetric (i.e., for $q=37$,
$\sigma(33\omega)\simeq\sigma(41\omega)$, etc.) and upon variation
of $q$ it shifts but remains almost invariant. }
\end{figure}

The effect at the single-atom level, however, is less clear. It
has been shown that the XUV photons control the timing of
ionization, and preferentially select certain quantum paths of the
electron \cite{K. J. Schafer+M. B. Gaarde}. While this effect may
lead to the enhancement of the low-order harmonics in the plateau,
it can't account for the large enhancement in the cutoff and
beyond (Fig.\ref{fig1}). A 3-step-model classical analysis of HHG
suggests that the contribution of the XUV field to the kinetic
energy of the returning electron is negligible. The kinetic energy
of a classical free electron of charge $e$ and mass $m$, driven by
a linearly-polarized strong IR fundamental field of frequency
$\omega$, amplitude $\varepsilon^{in}_{1}$ and polarization
$\mathbf{e_{k}}$
($\mathbf{E_{1}}(t)=\mathbf{e_{k}}\varepsilon^{in}_{1}cos(\omega
t)$) is $E_{k}(t)=\frac{p^{2}(t)}{2m}$.
$p(t)=\frac{e\varepsilon^{in}_{1}}{\omega} [sin(\omega
t)-sin(\omega t_{i})]$ is the momentum of the electron, and it has
been assumed that the electron is freed at time $t_{i}$ with zero
momentum. The addition of a   weak harmonic XUV field of frequency
$q\omega$ (where $q$ is a large integer) and amplitude
$\varepsilon^{in}_{q}$
($\varepsilon^{in}_{q}<<\varepsilon^{in}_{1}$) with the same
polarization,
($\mathbf{E_{q}}(t)=\mathbf{e_{k}}\varepsilon^{in}_{q}cos(q\omega
t)$), adds a small correction to the momentum, which is
proportional to $\frac{\varepsilon^{in}_{q}}{q\omega}$. As a
result, the correction to the kinetic energy, which appears in the
form of two additional terms, proportional to
$\frac{\varepsilon^{in}_{q}}{q\omega}$ and
$(\frac{\varepsilon^{in}_{q}}{q\omega})^{2}$, is negligible. Thus,
the additional XUV field will not affect the electron trajectories
and will not contribute to their kinetic energy. For this reason
the relative phase between the two fields doesn't play a role in
the HGS, which is indeed verified in both classical analysis and
quantum mechanical simulations (a small $q$, however, will affect
the dynamics differently \cite{T. Pfeifer, Andiel}). In addition,
assigning the electron a non-zero initial momentum to account for
the photoelectric effect, will not increase its kinetic energy
upon recombination.

An illustrative TDSE simulation however (Fig.\ref{fig1}) shows an
enhancement of the cut-off harmonics and the harmonics $q \pm 2K$
($K$ integer) upon addition of a weak XUV field to the strong IR
field. Moreover, the HGS possesses certain symmetries: with
respect to its center at harmonic $q$, the distribution of
harmonics of the enhanced part of the spectrum (harmonics that
have been produced only due to the addition of the XUV field), is
symmetric with respect to $q$ and remains almost invariant upon
variation of $q$. This suggests that despite the fact that the
additional weak XUV field doesn't affect the electron
trajectories, it does affect the recombination process. As will be
shown later, the XUV field induces periodic modulations to the
remaining ground electronic state, with the same frequency as the
XUV field. The returning electronic wavepacket recombines with
this modulated ground state to emit new harmonics. The purpose of
this article is to reveal this mechanism which is responsible for
the amplification phenomena due to the inclusion of the weak XUV
field and to prove that the enhancement is a robust single atom
phenomenon. The mechanism could suggest new types of HHG
experiments. It is not limited to the description of the
self-occurring case in monochromatic HHG experiments where XUV
radiation, generated by the leading edge of the IR pulse,
co-propagates with the IR field to form a bichromatic driver field
in the last part of the medium (thus leading to the extension of
the cutoff energy in real experiments as compared to single-atom
calculations). In cases where the XUV field saturates, it might be
useful to add it externally. For example, He is known to produce
higher harmonics than Ar. Hence, the support of the HGS obtained
from Ar can be dramatically extended by shining the Ar with a high
harmonic obtained from He (which is absent in the Ar HGS) in
addition to the strong IR field.

In order to reveal the enhancement mechanism due to the inclusion
of the weak XUV field, we study the dynamics of a
single active electron in an atom described by the field-free
Hamiltonian $H_{0}(\mathbf {r})$ subjected to a long pulse of the IR
field $\mathbf{E_{1}}(t)$ in the length gauge and under the dipole
approximation. The long pulse evolves the system adiabatically
\cite{Avner Adiabatic} from the initial ground state of the
field-free Hamiltonian $|\phi_{1}(\mathbf{r})\rangle$ to a single
resonance Floquet eigenstate $|\psi^{(0)}_{1}(\mathbf{r},t)\rangle$
of Eq.\ref{eq3} \cite{N. Moiseyev and F. Weinhold} which describes
the entire dynamics of the system. A formalism of \textit{time
independent} perturbation theory is applicable since the time $t$ may be
treated as an additional coordinate \cite{H. Sambe,t-tp}. In the
following, all the parameters $m,m',n,n',M,K$ denote integers ($\in
\textbf{Z}$).

\begin{equation}
H^{(0)}_{F}(\mathbf{r},t)|\psi^{(0)}_{j,m}(\mathbf{r},t)\rangle=\varepsilon^{(0)}_{j,m}
|\psi^{(0)}_{j,m}(\mathbf{r},t)\rangle \label{eq3}
\end{equation}
where $H^{(0)}_{F}(\mathbf{r},t)=H_{0}(\mathbf{r})-e \mathbf{r}
\cdot \mathbf{E_{1}}(t)-i\hbar\frac{\partial}{\partial t}$ is the
Floquet Hamiltonian. The indices $(j,m)$ label the eigenstates $j$
within any given Brillouin zone $m$ and $\mathbf{r}$ describes the
internal degrees of freedom. The Floquet eigenfunctions of this
operator satisfy the c-product inner product
\cite{Reviewnimrod,Wilkinson} (written in the usual dirac notation)
$\langle
\psi^{(0)}_{j,m}(\textbf{r},t)|\psi^{(0)}_{j',m'}(\textbf{r},t)
\rangle_{\mathbf{r},t}=\delta_{jj'}\delta_{mm'}$ and form a complete
set. Floquet eigenfunctions which lie within the $m$-th Brillouin
zone may be defined as $|\psi^{(0)}_{j,m}(\mathbf{r},t)\rangle
\equiv |\psi^{(0)}_{j}(\mathbf{r},t)\rangle e^{i\omega m t}$ and
$\langle\psi^{(0)}_{j,m}(\mathbf{r},t)| \equiv \langle
\psi^{(0)}_{j}(\mathbf{r},t)| e^{-i\omega m t}$ with energies
$\varepsilon^{(0)}_{j,m} \equiv \varepsilon^{(0)}_{j} +m\hbar
\omega$. The ket and bra Floquet eigenfunctions are periodic with
period $T\equiv 2\pi/\omega$ and can therefore be decomposed as a
Fourier sum
$|\psi^{(0)}_{j}(\mathbf{r},t)\rangle=\sum_{n}|\varphi^{(0)}_{j,n}(\mathbf{r})\rangle
e^{i\omega n t}$ and
$\langle\psi^{(0)}_{j}(\mathbf{r},t)|=\sum_{n}\langle\varphi^{(0)*}_{j,n}(\mathbf{r})|e^{-i\omega
n t}$. Note that the Fourier components of the bra state are not
complex-conjugated \cite{Reviewnimrod}.

In order to calculate the HGS one may assume the Larmor
approximation \cite{Jackson} and analyze the time-dependent
acceleration expectation value $\mathbf{a}^{(0)}_{1}(t)\equiv
\frac{\partial^{2}}{\partial
t^{2}}\langle\psi^{(0)}_{1}(\mathbf{r},t)|\mathbf{r}|\psi^{(0)}_{1}(\mathbf{r},t)\rangle_{\mathbf{r}}$
which is proportional to the emitted field. The acceleration in
energy space is given by the Fourier transform
$\mathbf{a}^{(0)}_{1}(\Omega)=\frac{1}{T}\int_{0}^{T}dt\mathbf{a}^{(0)}_{1}(t)e^{-i\Omega
t}$. Exploring only frequencies which are integer multiples of
$\omega$ [$\Omega=M\omega$ ( $M\in \textbf{Z}$)], and using the property $\frac{1}{T}\int_{0}^{T}dt~e^{-i\omega n
t}=\delta_{n,0}$, the expression obtained is
$\mathbf{a}^{(0)}_{1}(M\omega)=-\omega^{2}M^{2}\sum_{n}
\langle\varphi^{(0)*}_{1,n}(\mathbf{r})|\mathbf{r}|
\varphi^{(0)}_{1,n+M}(\mathbf{r})\rangle_{\mathbf{r}}$. It can be shown to be
non-vanishing only for integer odd values of $M$, which is a
well known feature of monochromatic HHG \cite{Ofir+Vitali}.

Suppose the weak XUV field $\mathbf{E_{q}}(t)$ is added. A
new Floquet problem is obtained, which could be described by the Floquet
Hamiltonian $H^{NEW}_{F}(\mathbf{r},t)\equiv
H^{(0)}_{F}(\mathbf{r},t)+V(\mathbf{r},t)$, where the
additional term $V(\mathbf{r},t)=-e \mathbf{r} \cdot
\mathbf{E_{q}}(t)$ could be treated as a perturbation. Time-independent
1st-order perturbation theory may be used to get an approximate solution for
the Floquet Hamiltonian $H^{NEW}_{F}(\mathbf{r},t)$ as

\begin{equation}
|\psi^{NEW}_{1}(\mathbf{r},t)\rangle=|\psi^{(0)}_{1}(\mathbf{r},t)\rangle
+ \sum_{(j',m')\neq (1,0)}
c^{j',m'}_{1}(q)|\psi^{(0)}_{j'}(\mathbf{r},t)\rangle e^{i \omega
m' t} \label{eq5}
\end{equation}
where the coefficients $c^{j',m'}_{1}(q)$ are given by

\begin{widetext}
\begin{equation}
c^{j',m'}_{1}(q)=-\frac{1}{2}e
\varepsilon^{in}_{q}\mathbf{e_{k}}\cdot\sum_{n}
\frac{\langle\varphi^{(0)*}_{j',n}(\mathbf{r})|\mathbf{r}|
\varphi^{(0)}_{1,n+m'-q}(\mathbf{r})\rangle_{\mathbf{r}}+\langle\varphi^{(0)*}_{j',n}(\mathbf{r})|\mathbf{r}|
\varphi^{(0)}_{1,n+m'+q}(\mathbf{r})\rangle_{\mathbf{r}}
}{\varepsilon^{(0)}_{1}-\varepsilon^{(0)}_{j'}-m'\hbar\omega}.
\label{eq6}
\end{equation}
\end{widetext}
Using this solution the time dependent
acceleration expectation value $\mathbf{a}^{NEW}_{1}(t)\equiv
\frac{\partial^{2}}{\partial
t^{2}}\langle\psi^{NEW}_{1}(\mathbf{r},t)|\mathbf{r}|\psi^{NEW}_{1}(\mathbf{r},t)\rangle_{\mathbf{r}}$ can be calculated. Keeping terms up to first order in $\varepsilon^{in}_{q}$ the following expression for the acceleration in the frequency domain is obtained:

\begin{widetext}
\begin{eqnarray}
\nonumber && \mathbf{a}^{NEW}_{1}(M\omega)= \mathbf{a}^{(0)}_{1}(M\omega) \\
&& -\omega^{2}M^{2} \sum_{(j',m')\neq (1,0)}\sum_{n}
[c^{j',m'}_{1}(q)\langle\varphi^{(0)*}_{1,n}(\mathbf{r})|\mathbf{r}|
\varphi^{(0)}_{j',n-m'+M}(\mathbf{r})\rangle_{\mathbf{r}}+c^{j',m'*}_{1}(q)\langle\varphi^{(0)*}_{j',n-m'-M}(\mathbf{r})|\mathbf{r}|
\varphi^{(0)}_{1,n}(\mathbf{r})\rangle_{\mathbf{r}}] \label{eq7}
\end{eqnarray}
\end{widetext}
This is the expression for the emitted HHG field. The HGS
($\sigma(M\omega)\equiv|\mathbf{a}^{NEW}_{1}(M\omega)|^{2}$) has
the same features as those presented in Fig.\ref{fig1} (see \cite{Avner Extended}). The weak
perturbative XUV field shifts the HGS beyond the cutoff obtained
by the IR field alone. In the Floquet formalism presented here \textit{the
origin of the HHG enhancement phenomena lies in the interferences
between the ground and excited Floquet states}. The HGS is
modified due to the dipole cross-terms introduced by the weak XUV
field.

The features in the HGS could also be explained in terms of the
re-collision model. It was shown that the additional weak XUV
field doesn't affect the electron trajectories, i.e., doesn't
modify the kinetic energy of the re-colliding electron. According
to the findings of the numerical simulation it must however affect
the recombination process. To see this we turn into the
semiclassical re-collision model \cite{P. B. Corkum} where the
electronic wavefunction at the event of recombination could be
described as a sum of the following continuum and bound parts.
Under the strong field approximation the returning continuum part
in the direction of the polarization $\mathbf{e_{k}}$ (which we
take as the x-direction from now on for simplicity) is a
superposition of plane waves
$\psi_{c}^{\parallel}(x,t)=\frac{1}{\sqrt{2\pi}}\int_{-\infty}^{\infty}dk
\tilde{\psi_{c}}(k,t)e^{i[kx-\frac{E_{k}}{\hbar} t]}$ where
$\mathbf{k}=k\mathbf{e_{x}}$ ($k=|\mathbf{k}|$) is the momentum of
the electron, $E_{k}\equiv \frac{\hbar^{2}k^{2}}{2m}$ is the usual
dispersion relation and $\tilde{\psi_{c}}(k,t)$ are expansion
coefficients which weakly depend on time. It is assumed that the
continuum wavepacket $\psi_{c}(\mathbf{r},t)$ is separable in the
x-coordinate and the 2 other lateral coordinates such that
$\psi_{c}(\mathbf{r},t)=\psi_{c}^{\parallel}(x,t)\psi_{c}^{\perp}(y,z,t)$.
It is assumed that the ground state is only slightly depleted
during the tunnel-ionization and that due to the ac-Stark effect
the electron adiabatically follows the instantaneous ground state
of the potential which is periodically modified by the IR and XUV
fields. Since the ac-Stark corrections to the instantaneous energy
and wavefunction are small for normal field intensities, the
instantaneous ground state could be approximated as
$\psi_{b}(\mathbf{r},t)\cong \phi_{1}(x+\varepsilon_{1}^{out}
cos(\omega t)+\varepsilon_{q}^{out} cos(q\omega t),y,z)
e^{+\frac{i}{\hbar}I_{p}t}$ (where $I_{p}>0$ and
$\phi_{1}(\mathbf{r})$ are the field-free ground state eigenvalue
and eigenstate, respectively). Note that $\psi_{b}(\mathbf{r},t)$
approximately describes the resonance Floquet state
$|\psi^{NEW}_{1}(\mathbf{r},t)\rangle$. The quiver amplitudes
$\varepsilon_{1}^{out}, \varepsilon_{q}^{out}$ of the spatial
oscillations of the ground state are of the order of
$\varepsilon_{q}^{out}=\frac{\varepsilon_{q}^{in}}{q^{2}\omega^{2}}$,
i.e., a tiny fraction of a Bohr radius for normal laser
intensities and/or large values of $q$. The bound part may
therefore be expanded in a Taylor serie as
$\psi_{b}(\mathbf{r},t)\cong e^{+\frac{i}{\hbar}I_{p}t}
\{\phi_{1}(\mathbf{r})+[\varepsilon_{1}^{out} cos(\omega
t)+\varepsilon_{q}^{out} cos(q\omega t)]\frac{\partial}{\partial
x}\phi_{1}(\mathbf{r})\}$. Using the total wavefunction at the
event of recombination
$\Psi(\mathbf{r},t)=\psi_{b}(\mathbf{r},t)+\psi_{c}(\mathbf{r},t)$,
the time-dependent acceleration expectation value
$\mathbf{a}(t)\equiv\frac{1}{m}\langle \Psi(\mathbf{r},t)|-\nabla
V_{0}(\mathbf{r})|\Psi(\mathbf{r},t) \rangle_{\mathbf{r}}$ could
be calculated, where $V_{0}(\mathbf{r})$ is the field-free
potential. The dominant terms that are responsible for the
emission of radiation at frequencies other than the incident
frequencies $\omega$ and $q\omega$ are the bound-continuum terms
$\mathbf{a}(t)\simeq 2\Re \langle \psi_{b}(\mathbf{r},t)|-\nabla
V_{0}(\mathbf{r})|\psi_{c}(\mathbf{r},t) \rangle_{\mathbf{r}}$.
After some algebra it could be realized that the acceleration is
composed of oscillating terms of the form

\begin{widetext}
\begin{eqnarray}
\mathbf{a}(t)\simeq -2\Re
\frac{1}{\sqrt{2\pi}}\int_{-\infty}^{\infty}dk
[\tilde{\tilde{\psi}}_{IR}(k)
e^{-\frac{i}{\hbar}[E_{k}+I_{p}]t}+\tilde{\tilde{\psi}}_{XUV}(k)
\varepsilon_{1}^{out} e^{-\frac{i}{\hbar}[E_{k}+I_{p}\pm \hbar
\omega]t}+\tilde{\tilde{\psi}}_{XUV}(k) \varepsilon_{q}^{out}
e^{-\frac{i}{\hbar}[E_{k}+I_{p}\pm q\hbar \omega]t}] \label{eq8}
\end{eqnarray}
\end{widetext}
where $\tilde{\tilde{\psi}}_{IR}(k)\equiv
\frac{1}{m}\tilde{\psi_{c}}(k)\int_{-\infty}^{\infty} d^{3}r
\phi_{1}(\mathbf{r})\nabla V_{0}(\mathbf{r})
\psi_{c}^{\perp}(y,z,t)e^{ikx}$ and
$\tilde{\tilde{\psi}}_{XUV}(k)\equiv
\frac{1}{2m}\tilde{\psi_{c}}(k)\int_{-\infty}^{\infty} d^{3}r
\frac{\partial \phi_{1}(\mathbf{r})}{\partial x}\nabla
V_{0}(\mathbf{r}) \psi_{c}^{\perp}(y,z,t)e^{ikx}$ and the $\pm$
sign in each of the last two terms stands for summation over
two-terms each. The emitted field in a single re-collision event
is composed of a continuum of these frequencies.

It is therefore seen that despite of their small magnitude, the
periodic time-dependent modulations to the ground electronic state
induced by the XUV weak field of frequency $q\omega$ are
responsible for the appearance of the new harmonics around $q$ in
the HGS via recombination with the returning electronic
wavepacket. Each electron trajectory (plane wave) with kinetic
energy $E_{k}$, recombines with the nucleus to emit, with equal
probabilities, one of three possible photons with energies:
$I_{p}+E_{k}$, $q\hbar\omega+I_{p}+E_{k}$ or
$q\hbar\omega-(I_{p}+E_{k})$. The HGS in the presence of the IR
field alone $\hbar\Omega=I_{p}+E_{k}$ is now shifted by the energy
of the XUV photon $\hbar q\omega$, and new harmonics are also
formed, such that their distribution about the center $q$ is
symmetric. Also, with respect to the center $q$, the distribution
of the XUV-formed harmonics, is invariant to a change in the
energy of the XUV photon $\hbar q\omega$, since these harmonics
are "born" from the same set of electron trajectories which are
characteristic of the IR field alone. When each single
re-collision event is repeated every half cycle of the IR field,
integer harmonics $q\pm 2K$ are obtained in the HGS. To see this, note that in two consecutive re-collision events at times $t_{r}$ and $t_{r}+T/2$ the following symmetry holds: $\tilde{\psi_{c}}(k,t_{r}+T/2)=\tilde{\psi_{c}}(-k,t_{r})$. Consequently, since $V_{0}(\mathbf{r})$ and $\phi_{1}(\mathbf{r})$ are symmetric functions for atoms (and $\frac{\partial \phi_{1}(\mathbf{r})}{\partial x}$ is antisymmetric), the following symmetry holds $\tilde{\tilde{\psi}}_{IR}(k,t_{r}+T/2)=-\tilde{\tilde{\psi}}_{IR}(-k,t_{r})$. The acceleration which results from the IR field therefore switches signs between subsequent re-collision events, which is the origin of the odd-selection rules. However, the behavior of the coefficients resulting from the addition of the XUV filed is different $\tilde{\tilde{\psi}}_{XUV}(k,t_{r}+T/2)=+\tilde{\tilde{\psi}}_{XUV}(-k,t_{r})$.  The acceleration which results from the additional XUV field doesn't switch signs between subsequent re-collision events and therefore yields even harmonics around $q$.

\begin{figure}[ht]
\hbox{\includegraphics[width=2.8in,height=2.8in]{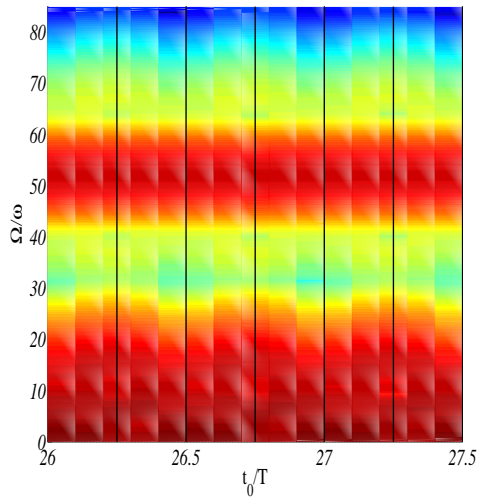}}
\caption{\label{fig2}\small (color online) Top view (red color-
high intensity, blue color- low intensity) of the absolute square
of the Gabor-tansformed acceleration expectation value (
$\frac{1}{50T}\int_{0}^{50T}a(t)e^{-\frac{(t-t_{0})^{2}}{\tau^{2}}}e^{-i\Omega
t}$, $\tau=0.1T$) of the quantum mechanical simulation described
in Fig.\ref{fig1} for $q=52$, as function of $t_{0}$ and $\Omega$.
}
\end{figure}

The above suggestion could be verified by plotting the
time-frequency distribution of high harmonics (Fig.\ref{fig2})
obtained from the time-dependent acceleration expectation value
whose spectra is given in Fig.\ref{fig1} for $q=52$. In accordance
with the classical re-collision model, different harmonics are
emitted repeatedly every half cycle, with the IR cut-off harmonic
(the 15th harmonic) emitted at times $\sim 0.2T+K\cdot 0.5T$. At
those instants, also the 38th and 66th harmonics, which are
produced by the most energetic IR trajectory, are emitted. Each electron trajectory in general, which under the IR field alone produces an harmonic $\Omega$, generates upon the addition of the XUV field, two duplicated new harmonics with energies $q\hbar\omega+\hbar\Omega,
q\hbar\omega-\hbar\Omega$, and similar properties. For example the
harmonics of orders $38-48$ and $56-66$ have a "plateau" character
(constant intensity), like the plateau harmonics $5-15$. Moreover,
as Eq.\ref{eq8} predicts, the intensity ratio of the enhanced-plateau
harmonics and the IR-plateau harmonics should be (for any of the values of $q$ given in
Fig.\ref{fig1})
$(\varepsilon_{q}^{out})^{2}=(\frac{\varepsilon_{q}^{in}}{q^{2}\omega^{2}})^{2}\simeq
10^{-10}$, in agreement with the results of Fig.\ref{fig1}.

In conclusion, we have shown that the addition of a weak XUV
harmonic field to a strong IR field leads to the extension of the
cut-off in the HGS. The results of the quantum analytical
expressions, quantum numerical simulations and classical arguments
suggest that the enhancement is a single-atom phenomena. The seed
XUV field modulates the ground state and affects the recombination
process of all returning trajectories, and leads to the generation
of new harmonics with structure well related to the HGS in the
presence of the IR field alone. This amplification mechanism for
the generation of high-order harmonics might be used to enhance
the yield of harmonics in HHG experiments.

This work was supported in part by the by the Israel Science
Foundation and by the Fund of promotion of research at the
Technion.

\bibliographystyle{plain}

\end{document}